# Solution of the Newtonian plane Couette flow with dynamic wall slip using machine-learning methods

**Georgia Foutsitzi[a], Nikolaos Antoniadis[a] and Georgios C. Georgiou[b*]**

[a]Department of Informatics and Telecommunications, University of Ioannina, Arta, GR-47100, Greece

[b]Department of Mathematics and Statistics, University of Cyprus, P.O. Box 20537, 1678, Nicosia, Cyprus

[*]Corresponding author: georgios@ucy.ac.cy

**Abstract**

This study presents a comparative investigation of Physics-Informed Neural Networks (PINNs) and data-driven Deep Operator Networks (DeepONets) for predicting the evolution of plane Newtonian Couette flow with dynamic wall slip. While traditional numerical methods, such as the Crank-Nicolson scheme, offer high accuracy, their computational demand poses challenges in real-time applications. To address this, we first implement a PINN framework to solve the governing equations for specific physical parameters. Subsequently, we develop a data-driven DeepONet, trained on high-fidelity numerical data, to learn the continuous solution operator across a broad range of slip boundary conditions and upper wall velocities. Our results indicate that while the PINN achieved superior point-wise precision with a relative $L_2$ error of $0.083\%$, it remains constrained by the requirement for instance-specific retraining. In contrast, the DeepONet demonstrates robust generalization on unseen and out-of-distribution signals with a mean relative error of $0.36\%$ and $0.88\%$, respectively. Most notably, it provides near-instantaneous inference, achieving a speedup factor of approximately $540\times$ over the numerical solver and $30.5\times$ over the PINN. This work demonstrates the synergy between physics-based and data-driven architectures and establishes DeepONet as a highly efficient surrogate model for rapid parametric exploration and real-time fluid dynamics forecasting.

## 1. Introduction

The classical no-slip assumption that fluid particles stick to solid boundaries has been conveniently employed in theoretical and computational fluid mechanics. However, it is well established that the no-slip condition is violated in many flows of industrial interest, as well as in viscometric flows, and in microfluidics applications [1–3] and may be exhibited not only by complex fluids but also by Newtonian ones [4]. It is well documented that wall slip may affect the stability of polymer-melt extrusion, enhancing or in certain cases, suppressing instability [1,3]. It may also lead to inaccurate rheological parameter estimates in viscometric flows [5] and is routinely exhibited by Newtonian fluids in micro- and nano-fluidic devices [4].

The various proposed mechanisms of wall slip and the different phenomenological models are discussed in the reviews of Hatzikiriakos [2,6] and Malkin and Patlazhan [3].

Navier [7] was the first to propose a linear slip law expressing the slip velocity, i.e., the relative velocity of the adjacent fluid particles to that of the wall (which may be moving), $v_s$, as a function of the (local) wall shear stress, $\tau_s$,

$$v_s = \frac{1}{\beta}\tau_s \tag{1}$$

where $\beta$ is the slip coefficient. This varies with the temperature and the pressure [1,3] and depends on the fluid properties (e.g., molecular weight and distribution) and the properties of the wall-fluid interface. When $\beta \to \infty$ the standard no-slip condition is obtained ($v_s = 0$). Nonlinear extensions of the Navier slip law (1), e.g., the power-law slip model, have been proposed [2,3]. Equation (1) and these extensions are static, i.e., they work well at steady state but fail to capture any memory effects, such as transient overshoots and slip velocity relaxation. Pearson and Petrie [8] generalized Eq. (1) by adding a time-derivative term with a slip relaxation parameter, λ, resulting in a 'retarted' slip law,

$$v_s + \lambda\frac{\partial v_s}{\partial t} = \frac{1}{\beta}\tau_s, \tag{2}$$

where $t$ is the time. Equation (2) reduces to the static Navier-slip law (1) when $\lambda = 0$ or in steady-state flow. As $\lambda$ increases, memory effects become significant and the evolution of the slip velocity lags variations in wall shear stress. Memory effects have been observed in experiments on molten polymer [9,10]. Dynamic wall slip has also been reported in the molecular dynamic simulations of Thalakkottor and Mohseni [11] for both gas and simple-liquid flows near oscillating walls. Kazatchkov and Hatzikirakos [12] and Malkin and Patlazhan [3] pointed out that retarted wall slip becomes relevant at high shear rates. Since reaching a steady state in the presence of dynamic wall slip may take very long times, special attention is required, to obtain reliable estimates of the rheological parameters in transient rheometric experiments.

The implications of dynamic wall slip have been demonstrated in various recent analytical and numerical studies on transient one- and two-dimensional Newtonian Couette and Poiseuille flows [13–15]. The central result, in all flow configurations is that wall slip generally damps the flow development and this damping is enhanced as the slip relaxation parameter is increased, in good agreement with experimental observations [9,10]. In particular, the plane Couette flow is a prototype viscometric flow, but it is also widely used to study viscous shear-driven flow dynamics and wall–fluid interactions [16]. Its simplicity allows rigorous theoretical analysis while capturing essential physical mechanisms relevant to rheological and microfluidic applications.

Although classical numerical methods, such as finite-difference, finite-element, and Crank–Nicolson schemes, can deliver accurate and stable solutions for such problems, their computational cost increases significantly when repeated simulations are required. This limitation becomes especially pronounced in parametric studies, uncertainty quantification, optimization, and real-time flow control, where the governing equations must be solved repeatedly for varying boundary conditions and material parameters. The challenge is further exacerbated when the wall motion is time-dependent and when slip takes place, introducing additional parameters, such as the slip coefficient and the slip relaxation parameter [2,17].

In recent years, scientific machine learning (SciML) has emerged as a promising alternative to conventional CFD solvers, offering data-driven and physics-aware approaches for modeling, prediction, and control of complex flow systems [18,19]. By leveraging neural networks to learn latent representations of flow dynamics, SciML methods enable rapid evaluation once trained, often at a fraction of the computational cost of traditional solvers. Comprehensive reviews have documented the increasing adoption of machine learning in fluid mechanics for state estimation, flow prediction, optimization, and active control, highlighting its adaptability across a wide range of engineering applications [20–23].

Among SciML approaches, Physics-Informed Neural Networks (PINNs) have gained considerable attention as a mesh-free framework for solving PDEs [24]. PINNs are a class of neural networks that integrate physical laws—typically expressed as partial differential equations (PDEs) or other governing equations—directly into the training process of deep learning models. Unlike traditional neural networks that rely solely on data, PINNs embed these physical principles as constraints in the loss function, ensuring that the learned solutions adhere to both the available data and the underlying physical laws, such as conservation laws, boundary conditions, or dynamic equations. This approach allows PINNs to achieve high-precision predictions of unknown physical quantities even when data is sparse or noisy, and they are particularly useful for solving forward and inverse problems in scientific and engineering domains without the need for mesh generation or large labeled datasets [25–27]. However, a key limitation of PINNs is their lack of generalization across parametric variations: each new boundary condition or parameter set typically requires retraining or fine-tuning, which limits their scalability for real-time or multi-query applications.

To address this limitation, recent research has increasingly focused on neural operator learning, which aims to approximate mappings between infinite-dimensional function spaces rather than pointwise solutions. Neural operators are specifically designed to approximate solution operators of PDEs, making them well suited for fluid dynamics problems where both inputs and outputs are continuous fields rather than finite-dimensional vectors [28–30]. Within this class of models, Deep Operator Networks (DeepONets) [29,31] have emerged as a powerful and theoretically grounded framework for learning nonlinear operators. DeepONets exhibit discretization invariance and strong generalization capabilities, making them particularly well suited for parametric PDEs with variable boundary and initial conditions.

Further advances have incorporated physical constraints into operator learning frameworks, leading to physics-informed or physics-aware DeepONets. These hybrid approaches combine the generalization capability of operator learning with the robustness and physical consistency of physics-informed training, significantly reducing data requirements while improving extrapolation performance [32–34]. Recent studies have demonstrated the effectiveness of such models for heat conduction [33], advection–diffusion–reaction systems [35], and other parametric PDEs under complex and time-dependent boundary conditions, achieving orders-of-magnitude speedups compared to classical numerical solvers.

Motivated by these developments, the present work investigates the Newtonian plane Couette flow with linear dynamic wall slip, a problem that introduces both time-dependent boundary forcing and parametric slip behavior. While PINNs can accurately resolve individual flow configurations, their limited reusability across parameter space motivates the exploration of operator-based alternatives. In this study, we conduct a comparative analysis between a traditional PINN formulation and data-driven DeepONet architectures for the prediction of the velocity field.

Specifically, a PINN is employed to solve a representative moderate-slip case, serving as a physics-consistent reference solution, while a Balanced DeepONet is trained to learn the solution operator across a wide range of slip coefficients, slip relaxation parameters, and upper wall velocity functions. This comparison underscores the fundamental trade-off between physics-based optimization — precisely tailored to a single configuration — and operator learning approaches that enable rapid, generalized predictions across diverse settings.

By addressing dynamic boundary motion and slip relaxation within an operator learning framework, this work extends the applicability of DeepONets to unsteady viscous flows with complex wall dynamics. The results demonstrate the potential of neural operator models as reliable, real-time-capable alternatives to conventional CFD solvers for parametric flow problems, while also clarifying the complementary roles of PINNs and DeepONets in scientific machine learning for fluid mechanics.

The rest of the paper is organized as follows: Section 2 establishes theoretical foundations, recalling the main equations governing the Newtonian plane Couette flow with dynamic wall slip, followed by the formulations of Physics-Informed Neural Networks (PINNs) and Deep Operator Networks (DeepONets) for parametric boundary conditions. Section 3 presents the results and discussion, beginning with the application of the PINN model to specific Couette flow cases. This is followed by an evaluation of the data-driven DeepONet framework, i.e., its ability to map arbitrary boundary forcing functions to spatio-temporal velocity fields. Subsequently, a comparative analysis between PINNs and DeepONet is performed to highlight their respective strengths in terms of accuracy and computational efficiency. Finally, Section 4 summarizes the key findings of this study and discusses some ideas for future work.

## 2. Formulations

### 2.1 The Newtonian Plane Couette Flow with Dynamic Wall Slip

The physical problem considered is the time-dependent, incompressible Newtonian plane Couette flow between two infinite parallel plates separated by a distance $d$. Initially, at $t = 0$, the fluid is at rest everywhere. At time $t > 0$, the upper plate ($y = d$) begins moving in the $x$-direction with prescribed velocity $V_0 f(t)$, while the lower plate ($y = 0$) remains fixed. For a Newtonian fluid with constant density $\rho$ and dynamic viscosity $\mu$, the $x$-momentum equation reduces to the one-dimensional diffusion equation:

$$\rho \frac{\partial v}{\partial t} = \mu \frac{\partial^2 v}{\partial y^2}, \quad t \geq 0, \qquad 0 \leq y \leq d \tag{3}$$

Dynamic wall slip following Eq. (2) is assumed to occur only along the fixed plate, whereas the classical no-slip condition applies at the moving plate. Thus, the boundary conditions for the velocity $v(y,t)$ read:

$$v(0,t) + \lambda \frac{\partial v}{\partial t}(0,t) = \frac{\mu}{\beta} \frac{\partial v}{\partial y}(0,t), \qquad t > 0 \tag{4}$$

At the fixed lower plate, a linear dynamic slip law is employed, which generalizes the classical Navier slip condition by incorporating a relaxation term. This dynamic slip law accounts for the time-dependent response of the fluid-wall interface and takes the form [16]:

$$v(d,t) = V_0 f(t), t > 0 \tag{5}$$

Since the fluid is initially at rest, the initial condition is

$$v(y,0) = 0, \quad 0 \le y \le d. \tag{6}$$

To nondimensionalize the problem of Eqs. (3)-(6), we scale lengths by $d$, time by the diffusion time scale $d^2\rho/\mu$, and the velocity by the characteristic velocity $V_0$ of the moving plate. For simplicity, we retain the same symbols for the dimensionless variables. The dimensionless form of the initial-boundary value problem becomes:

$$\frac{\partial v}{\partial t} = \frac{\partial^2 v}{\partial y^2}, \ \ t > 0, \ \ 0 \le y \le 1 \tag{7}$$

subject to the boundary conditions:

$$v(0,t) + \Lambda_s \frac{\partial v}{\partial t}(0,t) = \frac{1}{B}\frac{\partial v}{\partial y}(0,t), \qquad t > 0 \tag{8}$$

$$v(1,t) = f(t), \qquad t > 0 \tag{9}$$

and the initial condition:

$$v(y,0) = 0, \quad 0 \le y \le 1. \tag{10}$$

Two dimensionless parameters emerge from the nondimensionalization process: the *slip number,*

$$B \equiv \frac{\beta d}{\mu} \tag{11}$$

and the *slip-relaxation number*

$$\Lambda_s \equiv \frac{\lambda \mu}{\rho d^2} \tag{12}$$

The slip number $B$ quantifies the ease of slip at the interface, with larger values indicating weaker slip (approaching no-slip as $B \to \infty$) and smaller values indicating stronger slip. The dimensionless slip relaxation number $\Lambda_s$ characterizes the temporal memory effect of the dynamic slip law, representing the ratio of the slip relaxation time scale to the diffusion time scale. When $\Lambda_s = 0$, the static Navier slip condition is recovered [2].

The initial-boundary value problem described by Eqs. (5)–(8) was solved by Abou Hasan et al. [16] using an efficient numerical method based on weighted-average finite differences (WAFD). Their study considered three distinct cases of motion for the moving plate at $y = 1$: constant speed, oscillatory speed, and single-period sinusoidal speed. The WAFD scheme with a weighting factor $\theta = 0.5$ corresponds to the well-known Crank–Nicolson (CN) implicit scheme. In the present work, we adopt this method to generate reference solutions, which serve as ground truth for both training and evaluation purposes.

## 2.2. Physics-Informed Neural Network (PINN)

Initially, the Newtonian plane Couette flow with linear dynamic wall slip is solved for the special case of $\Lambda_s = 0.5$, $B = 1$ (moderate slip) and $f(t) = \sin(4\pi t)$ (sinusoidal upper speed) using the Physics-Informed Neural Network (PINN) framework [24]. This approach embeds the governing physical laws directly into the training process of a neural network, allowing the simultaneous enforcement of the partial differential equation, boundary conditions, and initial condition without the need for labeled data.

The unknown velocity field $v(y,t)$ is approximated by a fully connected feed-forward neural network $v_\theta(y,t)$, parameterized by $\theta$, where $\theta$ denotes the set of all trainable parameters, including weights and biases, that are optimized during the training process. The network takes the spatial and temporal coordinates $(y,t)$ as inputs and outputs the predicted velocity (**Figure *1***). Multiple hidden layers with nonlinear activation functions (e.g., hyperbolic tangent) are employed to ensure sufficient representational capacity for smooth solutions of parabolic partial differential equation. Spatial and temporal derivatives required by the governing equation and boundary conditions are computed via automatic differentiation, which provides machine-precision accuracy and avoids numerical discretization errors.

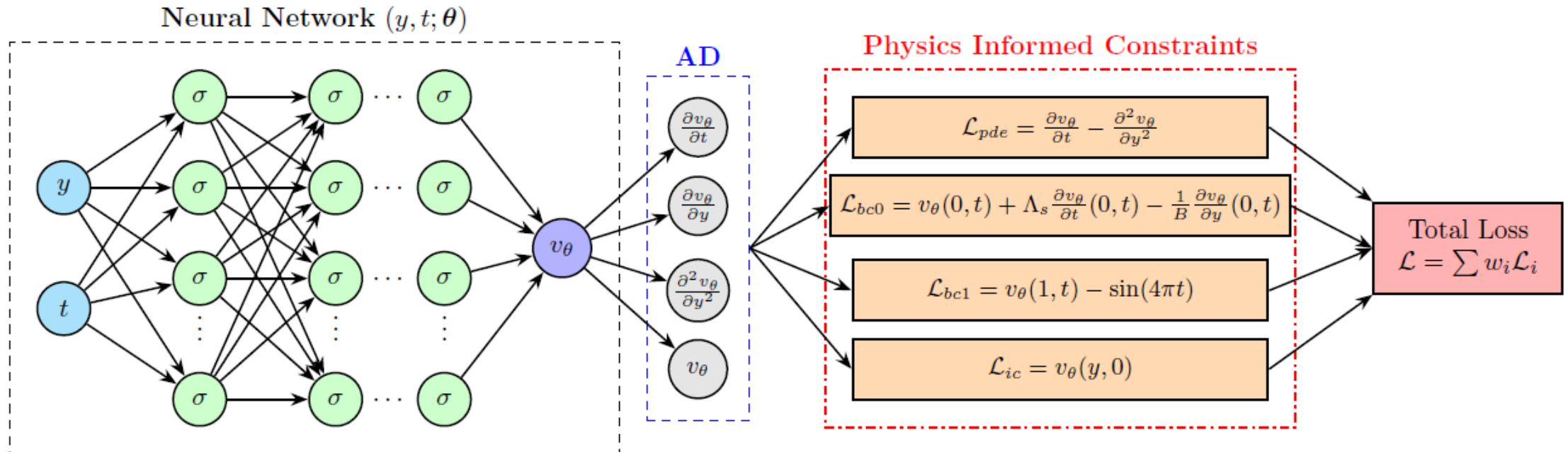

**Figure 1.** Schematic of the Physics-Informed Neural Network (PINN) architecture used to approximate the velocity field $v(y,t)$. The network takes the spatial coordinate $(y)$ and time $(t)$ as inputs and outputs the predicted velocity. Automatic differentiation is employed to compute the required derivatives, enabling enforcement of the governing diffusion equation, initial condition, and boundary conditions directly through the loss function.

The PINN model is trained by minimizing a composite loss function defined as [24]

$$\mathcal{L} = w_{pde}\mathcal{L}_{pde} + w_{bc0}\mathcal{L}_{bc0} + w_{bc1}\mathcal{L}_{bc1} + w_{ic}\mathcal{L}_{ic}, \tag{13}$$

where each term $\mathcal{L}_i$ enforces a specific physical constraint with a corresponding weight $w_i$.

The residual of the governing equation is given by

$$\mathcal{L}_{pde} = \frac{1}{N_{pde}} \sum_{i=1}^{N_{pde}} \left( \frac{\partial v_\theta}{\partial t}(y_i, t_i) - \frac{\partial^2 v_\theta}{\partial y^2}(y_i, t_i) \right)^2, \tag{14}$$

evaluated at $N_{pde}$ collocation points randomly sampled within the spatio-temporal domain.

The initial condition loss, defined as

$$\mathcal{L}_{ic} = \frac{1}{N_{ic}} \sum_{i=1}^{N_{ic}} (v_\theta(y_i, 0))^2, \tag{15}$$

quantifies the residuals of the initial condition on $N_{ic}$ points. The boundary condition loss consists of two parts:

$$\mathcal{L}_{bc0} = \frac{1}{N_{bc}} \sum_{i=1}^{N_{bc}} \left[ v_\theta(0, t_i) + \Lambda_s \frac{\partial v_\theta}{\partial t}(0, t_i) - \frac{1}{B} \frac{\partial v_\theta}{\partial y}(0, t_i) \right]^2 \tag{16}$$

and

$$\mathcal{L}_{bc1} = \frac{1}{N_{bc}} \sum_{i=1}^{N_{bc}} [v_\theta(1, t_i) - \sin(4\pi t_i)]^2, \tag{17}$$

where $N_{bc}$ is the total number of sampled points in time. Notice that all loss terms are a function of the network weight and bias parameters, denoted by $\theta = (w_i, b_i)$. These parameters are optimized using gradient-based optimization algorithms. An adaptive optimizer (Adam) [36] is employed first to achieve rapid convergence, followed by a quasi-Newton method (L-BFGS) [37] to further reduce the loss and improve solution accuracy.

The PINN methodology offers a flexible and robust framework for solving time-dependent flow problems with non-standard boundary conditions, such as linear dynamic wall slip. The resulting model provides a continuous, differentiable approximation of the velocity field over the entire space–time domain, making it well suited for parametric studies and further theoretical analysis.

### 2.3 Deep Operator Network (DeepONet) Formulation for Parametric Boundary Conditions

DeepONet, originally proposed by Lu et al. [31], provides a powerful framework for learning nonlinear operators that map between infinite-dimensional function spaces. This work extends the DeepONet architecture to learn the solution operator for the parametric plane Couette flow problem described in section 2.1. Unlike traditional PINNs that solve a PDE for a given set of input parameters, DeepONet learns the mapping from multiple input parameters—including functions and scalars—to the corresponding velocity field solution. This capability is particularly valuable for the present problem, where the solution depends on the time-dependent upper plate velocity function $f(t)$, as well as the two dimensionless numbers, $B$ and $\Lambda_s$.

We define the parameter space $\mathcal{P}$ consisting of elements $p = (f, \Lambda_s, B)$, where $f \in \mathcal{F}$ is a function in an infinite-dimensional function space representing the upper plate velocity and $\Lambda_s$, $B$ are positive scalar parameters. For each such parameter configuration, there exists a unique solution $v = v(p)(y, t)$ to the initial-boundary value problem defined by Eqs. (7)-(10). Consequently, we can define the solution operator $\mathcal{G}: \mathcal{P} \to \mathcal{S}$, where $\mathcal{S}$ is the solution space of velocity fields, such that

$$\mathcal{G}(p)(y, t) = v(p)(y, t) \tag{18}$$

The DeepONet architecture is designed to approximate the operator $\mathcal{G}$ by taking the parameter configuration $p$ and the spatio-temporal coordinates $(y, t)$ as inputs and returning the predicted velocity $\hat{v}(y, t)$. The architecture, illustrated in **Figure *2***, comprises two main sub-networks: a branch network and a trunk network.

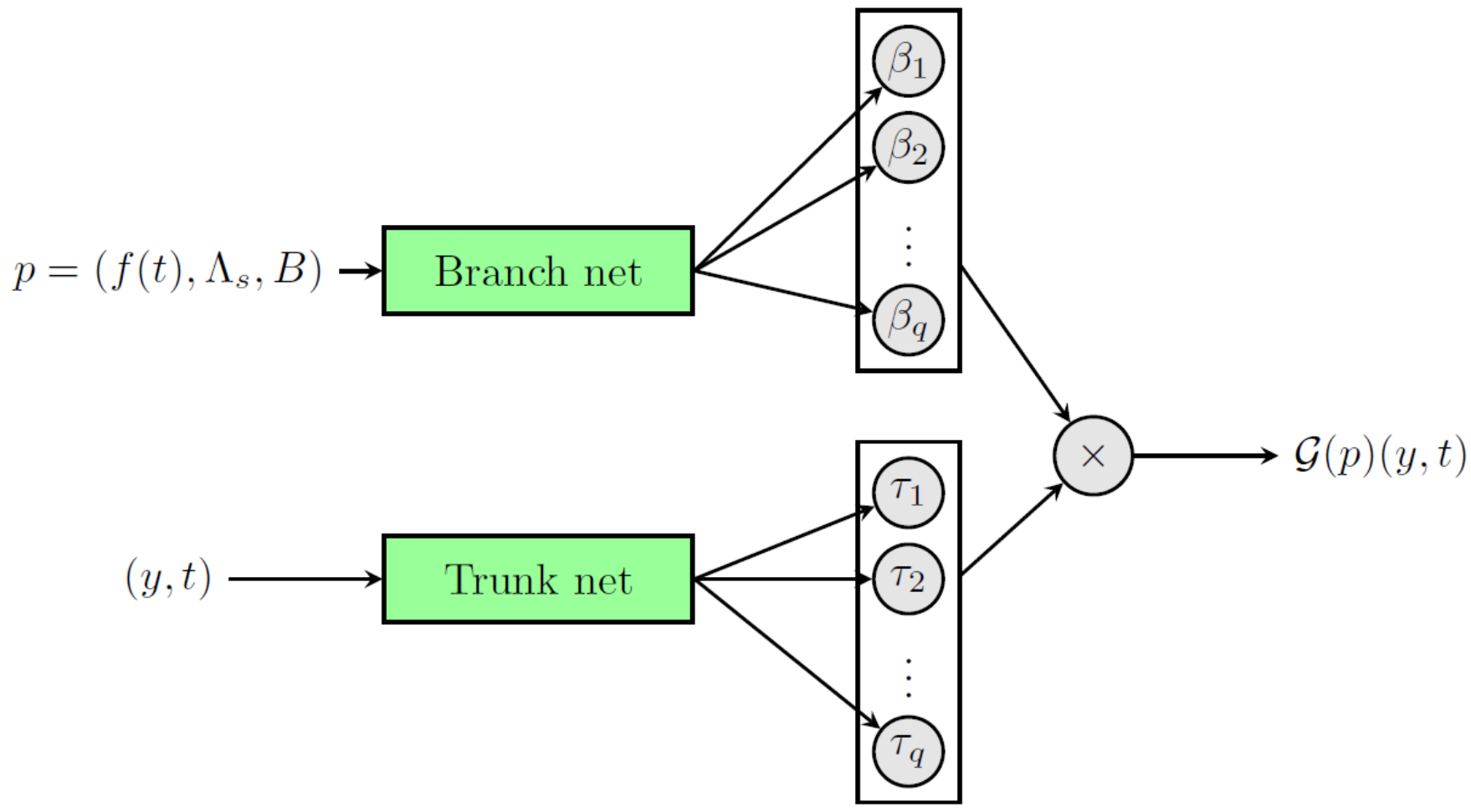

**Figure 2**. Schematic representation of the Deep Operator Network (DeepONet) architecture.

The *branch network* is responsible for processing all input parameters that characterize the specific problem instance. In our formulation, this includes:

1. *The function input* $f(t)$: In practice, the continuous function $f(t)$ is represented by its values at $m$ discrete sensor points in the time domain, denoted as $[f(t_1), f(t_2), \ldots, f(t_m)]$. These sensor points are typically chosen to cover the time interval of interest $[0, T]$.

2. *The scalar parameters*: The slip relaxation number $\Lambda_s$ and the slip number $B$ are appended to the discretized function values, forming a composite input vector of dimension $m + 2$:

$$\text{Branch input} = [f(t_1), f(t_2), \ldots, f(t_m), \Lambda_s, B]$$

The branch network, a fully-connected feed-forward neural network, processes this input and outputs a latent representation vector, $[\beta_1, \beta_2, \ldots, \beta_q]$, of dimension $q$ [31]. This vector can be interpreted as the coefficients for a set of learned basis functions that encode the influence of the input parameters on the solution.

The *trunk network* takes as input the spatio-temporal coordinates $(y, t)$ at which the solution is evaluated. For the plane Couette flow problem, $y \in [0,1]$ is the dimensionless distance from the fixed lower plate, and $t \in [0, T]$ is the dimensionless time. The trunk network, also a fully connected neural network, outputs another $q$-dimensional vector $[\tau_1, \tau_2, \ldots, \tau_q]$, which can be interpreted as a set of $q$ basis functions evaluated at $(y, t)$.

The final output of the DeepONet is the dot product of the branch and trunk network outputs, optionally with an added bias term:

$$\hat{\mathcal{G}}(p)(y,t) = \sum_{k=1}^{q} \beta_k(p) \cdot \tau_k(y,t) + b_0, \quad (19)$$

where $b_0$ is a trainable bias parameter. This elegant design enables the network to represent the solution as a linear combination of nonlinear basis functions (learned by the trunk network) with coefficients that are nonlinear functionals of the input parameters (learned by the branch network).

It is important to note that the unstacked DeepONet architecture employed here (see **Figure *1***), where the branch and trunk networks are separate and their outputs are combined via dot product, has been shown by Lu et al. [31] to achieve superior performance compared to stacked architectures while requiring fewer computational resources.

The DeepONet framework offers a significant advantage over traditional numerical solvers and standard PINNs for this problem. Once trained on a sufficiently representative set of parameter configurations, the network can rapidly predict the velocity field $v(y,t)$ for new combinations of the upper plate velocity function $f(t)$, the slip relaxation number $\Lambda_s$, and the slip number $B$. These predictions are obtained without retraining the model or re-solving the governing equations. Consequently, DeepONet provides an efficient surrogate for many-query applications, including uncertainty quantification, sensitivity analysis, and real-time control in microfluidic systems where wall-slip effects are important.

# 3. Results and Discussion

## 3.1 PINN

In this section, we use PINN model described in section 2.1 for solving the specific case of the Newtonian plane Couette flow (e.g. $B = 1, \Lambda_s = 0.5, f(t) = \sin(4\pi t)$). To obtain the velocity field $v(y,t)$, a neural network with four hidden layers, each consisting of 90 neurons, was employed. The hyperbolic tangent activation function was applied to all layers. The PINN model was trained on randomly sampled collocation points in the spatio-temporal domain $[0,1] \times [0,1]$. Specifically, $N_{pde} = 10^4$ interior points were used to enforce the PDE residual, while $N_{ic} = 3{,}000$ and $N_{bc} = 3{,}000$ points were used to impose the initial and boundary conditions, respectively. Training is based on the weighted MSE loss function (13) to balance the different terms. The residuals of the differential equation and boundary condition at $y = 0$ are weighted by $w_{pde} = w_{bc0} = 1$, while the errors associated with the initial and boundary condition at $y = 1$ are weighted by $w_{ic} = w_{bc1} = 10$. Different training configurations were tested, with the aforementioned weights giving the best results. Optimization was performed using a two-stage strategy: first, the Adam optimizer was applied for $10{,}000$ iterations at a learning rate of $10^{-4}$, followed by L-BFGS optimization with a maximum of $4{,}000$ iterations to refine the solution. This combined strategy ensures both training stability and accurate satisfaction of the physics constraints throughout the domain.

**Table 1**. Statistical performance of the proposed PINN architecture across five independent runs.

| Metric | Mean Value | Standard Deviation | Best Achievement |
|---|---|---|---|
| **Rel. $L_2$ Error (%)** | 0.083 | 0.006 | 0.075 |
| **Training Time (min)** | 46.83 | 7.3 | 38.20 |

The accuracy of the model was evaluated using the relative $L_2$ error metric

$$Rel\ L_2 = \frac{\left\| v_{pred} - v_{ref} \right\|_2}{\left\| v_{ref} \right\|_2} \tag{20}$$

which measures the global discrepancy between the predicted velocity field $v_{pred}$ and the reference solution $v_{ref}$ obtained from the CN solver.

To assess variability due to random initialization and sampling, five independent runs were repeated over different random seeds. For each run, the training time was recorded and the relative $L_2$ error was computed against the numerical reference solution (CN) on a uniform $200 \times 200$ evaluation grid. The results corresponding to the mean and standard deviation across all runs are given in **Table *1***. The results show consistently high accuracy, with a mean relative $L_2$ error of $0.083\%$ and a small standard deviation of $0.006\%$, indicating stable model performance. The best run achieved an error of $0.075\%$. The average training time was $46.83$ minutes, with some variability across runs due to differences in optimization convergence. The results show that the proposed PINN framework provides reliable and accurate solutions for the oscillatory Couette flow problem while maintaining consistent performance across multiple independent training runs.

Throughout the subsequent analyses, the results and the visualizations correspond to the representative model from each ensemble achieving median test set performance, ensuring that the reported efficiency and accuracy are indicative of the typical model behavior rather than an outlier.

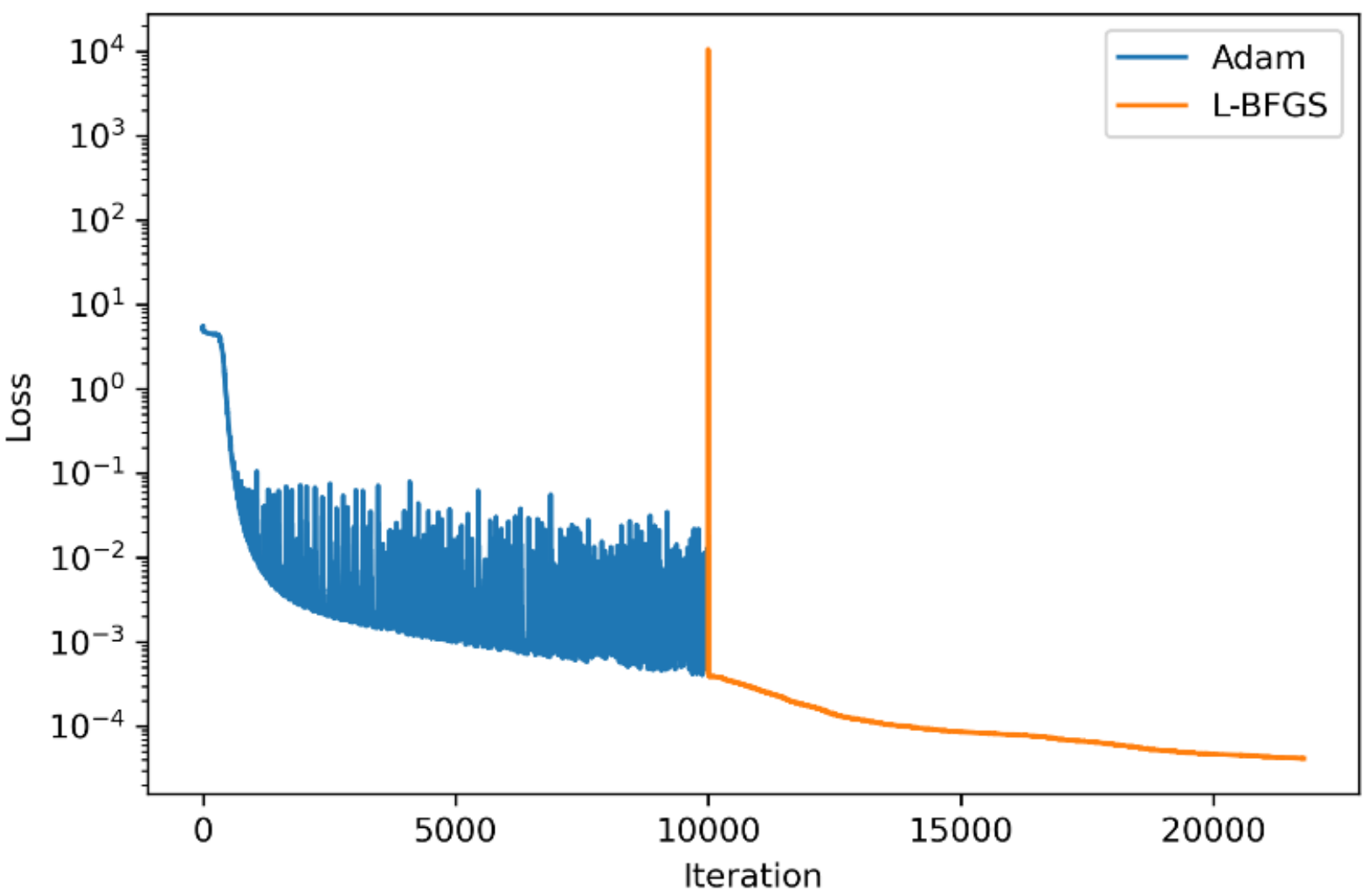

**Figure 3.** Training convergence of the PINN model showing the evolution of the total loss function. The initial phase (Adam optimizer) achieves rapid reduction of the loss, followed by refinement using the L-BFGS optimizer, which further decreases the residual by approximately one order of magnitude. The transition between optimizers is marked by a temporary increase in loss, after which stable convergence is achieved.

**Figure *3*** illustrates the training convergence of the proposed PINN model, reflecting the two-stage optimization strategy previously described. During the initial Adam stage, the loss function decreases rapidly by several orders of magnitude within the first few thousand iterations, indicating effective initial learning of the governing physics constraints. After approximately $10{,}000$ iterations, the loss gradually stabilizes around $10^{-4}$, suggesting that the optimizer approaches a plateau typical of first-order optimization methods. At this point, the L-BFGS optimizer is applied to further minimize the residual loss. A brief increase in the loss is observed at the transition between the two optimizers, which is commonly associated with the change in optimization strategy. Subsequently, L-BFGS significantly improves

convergence, reducing the loss by an additional order of magnitude and reaching values on the order of $10^{-5}$. This behavior is consistent with previous studies demonstrating that second-order optimization methods can substantially enhance the final accuracy of PINNs after the initial Adam training phase [38,39].

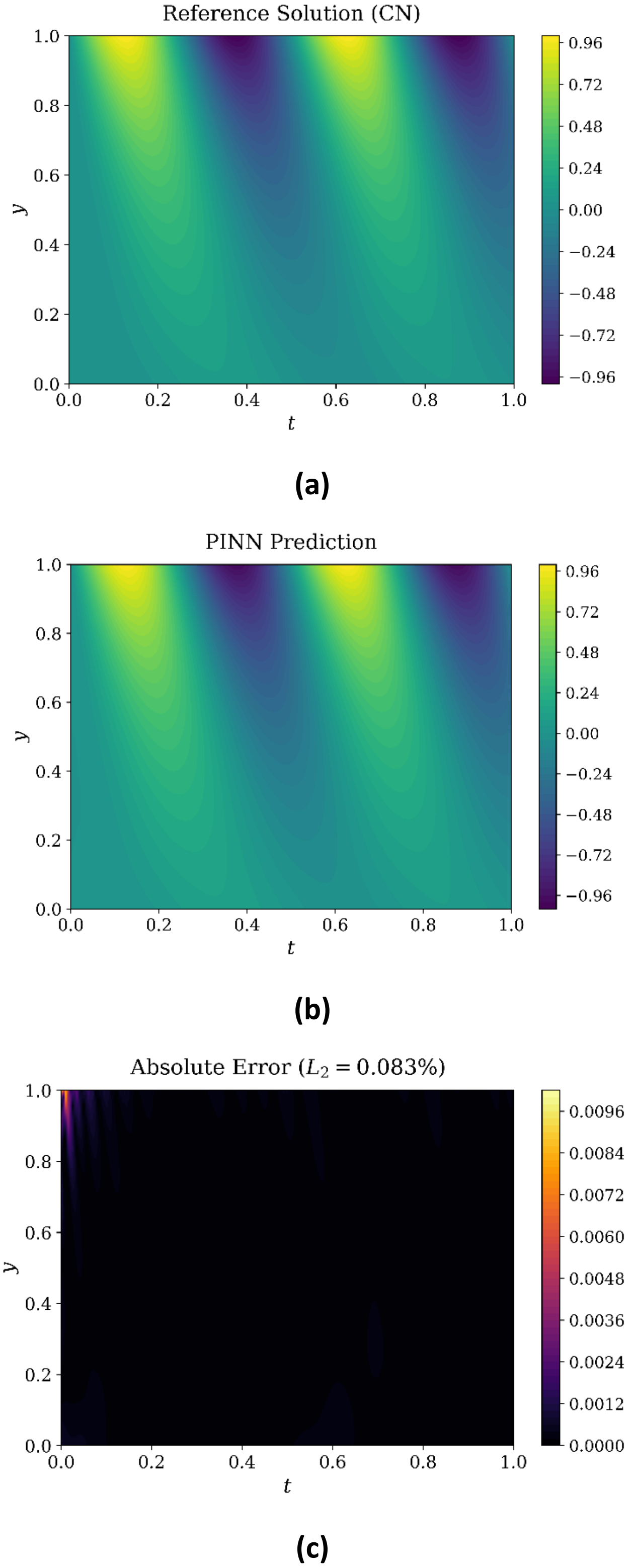

**Figure 4**. Results for oscillatory Couette flow with $B = 1$, $\Lambda_s = 0.5$ and $f(t) = \sin(4\pi t)$: (a) Numerical (CN) velocity; (b) PINN velocity prediction; (c) Absolute $L_2$ error.

**Figure 4** presents a comparison between the velocity field predicted by the proposed Physics-Informed Neural Network and the reference solution obtained using the CN numerical solver. The spatio-temporal evolution of the velocity field is shown over the entire computational domain $(y, t) \in [0,1] \times [0,1]$. As shown in the first two contour plots, the PINN prediction is in excellent agreement with the reference CN solution. The model accurately captures the characteristic oscillatory behavior induced by the time-dependent boundary condition imposed at the upper plate. In particular, the spatial decay of the oscillations from the upper boundary toward the lower wall is faithfully reproduced, demonstrating that the PINN successfully learns the underlying diffusion dynamics governing the system.

The absolute error field shown in **Figure 4**c, further illustrates the accuracy of the proposed approach. The error remains very low across most of the domain, with only slightly larger deviations observed near the upper boundary, where the oscillatory forcing introduces sharper gradients. Overall, the error magnitude is predominantly below $10^{-3}$, while the global relative $L_2$ error of the velocity field is $0.083\%$, indicating that the PINN provides an accurate approximation of the reference numerical solution.

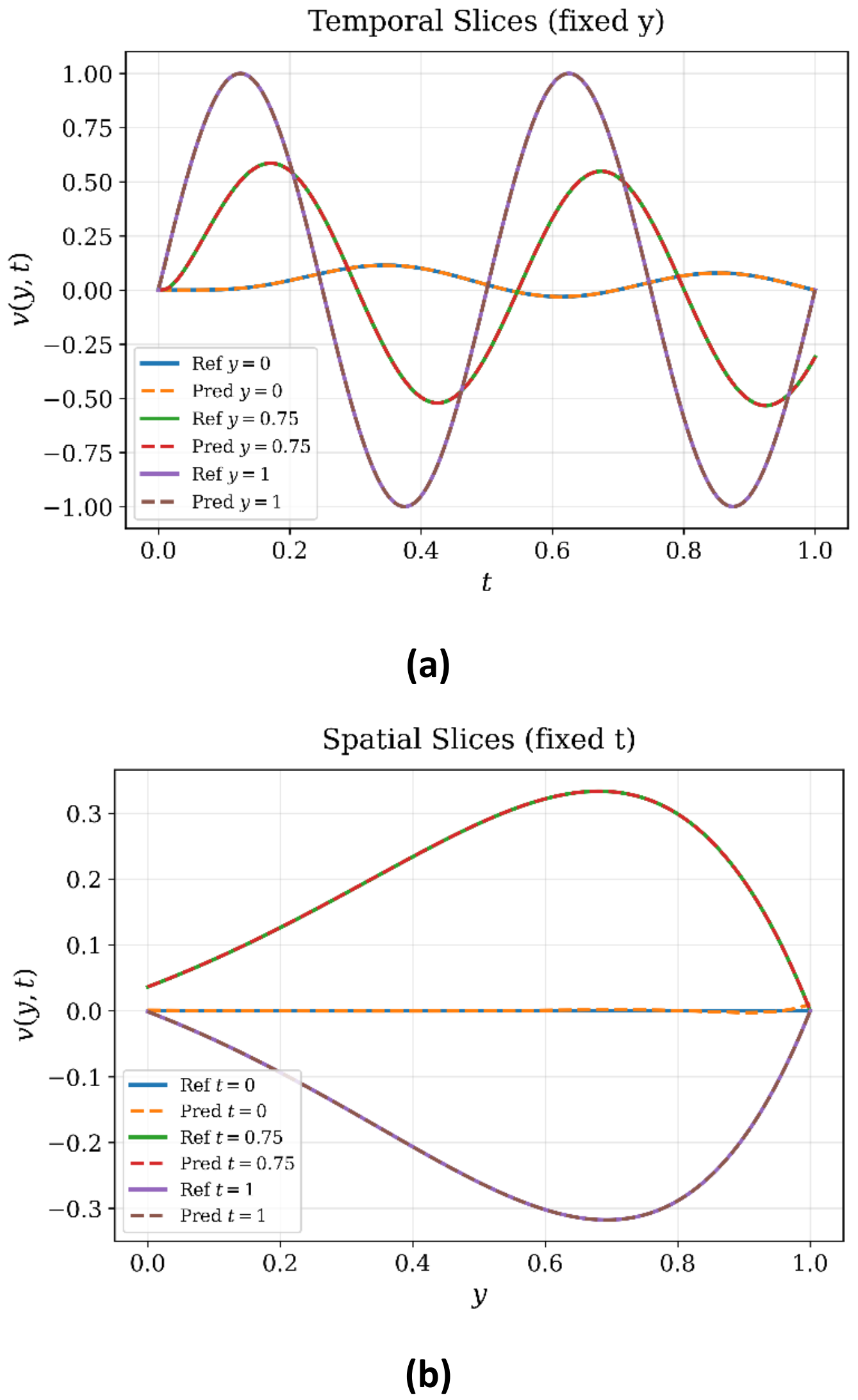

**Figure 5.** Comparison of the PINN predictions (dashed) with the reference solution (solid) with $B = 1$, $\Lambda_s = 0.5$ and $f(t) = \sin(4\pi t)$: (a) Temporal slices of the velocity field at fixed spatial positions $y$; (b) Spatial velocity profiles across the channel at selected time instances.

To further assess the accuracy of the proposed PINN model, additional comparisons are performed using the temporal and spatial slices of the velocity field that are shown in **Figure *5***. These slices provide a more detailed pointwise validation of the predicted solution against the reference numerical results.

For the temporal slices (fixed $y$), the velocity evolution over time is examined at selected spatial locations within the channel. The comparison shows that the PINN predictions closely match the oscillatory behavior of the reference solution. In particular, the amplitude and phase of the oscillations are accurately captured, demonstrating that the neural network successfully learns the time-dependent dynamics induced by the oscillatory boundary condition at the upper plate.

Similarly, spatial slices (fixed $t$) are analyzed to evaluate the velocity distribution across the channel at specific time instances. The results indicate excellent agreement between the PINN predictions and the reference solution throughout the entire spatial domain. The neural network accurately reproduces the smooth decay of the velocity profile from the oscillating upper boundary toward the stationary lower wall.

Overall, these results demonstrate that the proposed PINN framework accurately reconstructs the spatio-temporal velocity field of oscillatory Couette flow. The near-perfect agreement with the reference numerical solution—observed in both temporal and spatial slices—confirms the model's ability to capture the underlying physical dynamics with high fidelity, achieving minimal approximation error throughout the domain.

## 3.2 Data-driven DeepONet

In this section, we apply the data-driven neural operator framework introduced in Section 2.3 to the Couette flow problem. A Deep Neural Operator (DeepONet) is trained to learn the mapping from arbitrary boundary forcing functions and slip parameters to the corresponding spatio-temporal velocity field.

**Dataset Generation**

The dataset was generated by numerically solving the Couette flow problem using the Crank-Nicolson finite-difference method [16], accelerated via GPU computing with the CuPy library [40]. We produced a total of $10{,}000$ unique simulations to ensure a miscellaneous representation of the fluid's response. For each sample, the spatio-temporal velocity field $v(y,t)$ was computed on a uniform grid of $N = 199$ spatial points and $M = 199$ temporal steps (resulting in a $200 \times\ 200$ output matrix). The overall dataset was split to $80\%$ for training, $10\%$ for testing and $10\%$ for validation.

The boundary forcing function $f(t)$—representing the moving wall velocity—was modeled as a Gaussian Random Field (GRF) [41] with a Radial Basis Function (RBF) kernel with a length scale parameter $l =\ 0.2$, ensuring smooth yet complex temporal profiles. The forcing function $f(t)$ was discretized on the same temporal grid consisting of $200$ time points.

The parametric ranges for the slip number ($B \in\ [0.1, 100]$) and the slip relaxation number ($\Lambda_s \in\ [0, 0.5]$) were selected to encompass a wide spectrum of flow regimes, ranging from shear-dominated to pressure-driven conditions. This specific interval allows the DeepONet to learn the transition from no-slip boundary conditions to significant interfacial velocity jumps, ensuring the model's robustness across physically relevant fluid-solid interaction scenarios.

To enhance the model's robustness across different physical regimes, we employed a hybrid sampling strategy for the dimensionless parameters: the slip number $B$ was sampled logarithmically in the range $[0.1, 100]$ to cover three orders of magnitude of viscous dissipation, while the slip relaxation number $\Lambda_s$ was sampled linearly in $[0, 0.5]$. Notably, $20\%$ of the $\Lambda_s$ values were explicitly set to zero to include cases of pure Navier slip conditions. Each simulation was stored as a compressed .npz file containing the discretized forcing function, the governing parameters, and the full velocity evolution field for training and validation.

**Computational Implementation of DeepONet**

At the heart of the present methodology lies a DeepONet architecture specifically optimized to handle the high-dimensional mapping from the boundary forcing functions and slip parameters to the resulting fluid flow response. Implemented in the TensorFlow/Keras [42,43] framework, this architecture adopts a dual-pathway approach to decouple the input function representation from the spatio-temporal domain.

The branch network acts as a feature extractor for the input space, processing the boundary forcing function $f(t)$ alongside the physical parameters $\Lambda_s$ and $B$. Its topology consists of four hidden layers, each containing $192$ neurons. While the standard Rectified Linear Unit (ReLU) was initially considered, preliminary numerical experiments indicated that it resulted in slower convergence and reduced accuracy in capturing the operator's response. In contrast, the GELU [44] activation provides a smooth, non-monotonic response that proved superior for this specific task. Its continuous derivative properties are particularly advantageous for approximating physical operators that govern continuous phenomena, such as the diffusion of momentum in fluid flow, where maintaining a stable and consistent gradient flow is critical for the optimization process.

In addition, the Trunk network is responsible for processing the spatio-temporal coordinates $(y, t)$. In the present implementation, the Trunk's topology is carefully balanced with that of the Branch, featuring four hidden layers of $192$ neurons each. Aligning the network widths and setting the latent dimensionality to $q = 256$ ensure that the spatial basis functions generated by the trunk possess sufficient expressivity to represent the intricate features extracted by the branch. Again, the GELU activation function is employed throughout all layers to maintain architectural consistency and support the learning of smooth basis functions.

Choosing four layers for each sub-network strikes a balance between model capacity and generalization. Extensive preliminary trials indicated that this depth provides the necessary non-linear expressivity to capture the steep velocity gradients near the boundaries, especially in high slip number regimes, while simultaneously mitigating the risk of overfitting. This balanced depth-to-width ratio ensures that the network remains computationally efficient during both training and real-time inference, without compromising the spectral accuracy required for operator learning.

The final stage of the architecture involves an interaction layer where the two nets are merged. This is achieved via a dot product of their respective $q$ dimensional latent outputs, followed by the addition of a trainable bias to produce the final scalar velocity prediction $v$. This inner-product structure allows the model to efficiently synthesize the learned basis functions with the input-dependent coefficients, effectively reconstructing the solution of the partial differential equation across the entire parametric regime.

### The Operator Validation Framework

A fundamental strategy within the proposed methodology is the implementation of a specialized Operator Validation Callback, designed to provide a more robust assessment of the model's performance by monitoring its physical consistency rather than relying solely on point-wise Mean Squared Error (MSE) loss. While the training phase minimizes the standard MSE, the validation process is governed by the relative $L_2$ error metric defined by Eq. (20) which evaluates the entire operator's predictive accuracy. This metric is particularly significant as it offers heightened sensitivity to structural deviations in the velocity field, ensuring that the global characteristics of the flow are preserved. To balance computational efficiency with rigorous monitoring, this validation procedure is executed every $10$ epochs. During each execution, the model predicts the velocity field across the entire spatio-temporal grid for each validation sample. Given the high resolution of the grid, memory management is optimized by processing coordinate points in batches of $4{,}096$. Furthermore, this framework incorporates a preservation mechanism that tracks the minimum operator error achieved throughout the training process. By saving only the model state that attains a new global minimum in validation error, we ensure that the final optimized weights represent the most physically accurate representation of the underlying fluid dynamics.

### Training Hyperparameters

A carefully tuned set of hyperparameters led to the development of the final high-performance model. A constant learning rate of $10^{-4}$ was maintained throughout the duration of the training to facilitate a stable and deep exploration of the complex loss domain, preventing early convergence to suboptimal minima. The training process utilized a Monte Carlo sampling strategy, where $4{,}096$ unique spatio-temporal points were randomly selected per iteration. With $500$ steps per epoch, the model is exposed to a vast dataset of $2 \times 10^6$ distinct points in every training cycle. This high-density sampling ensures that the neural operator effectively learns the continuous nature of the velocity field across the entire parametric space of the upper wall velocity, and the slip and slip-relaxation numbers.

### Predictive Accuracy and Convergence

To ensure the robustness of the learning process, the DeepONet model was trained using five independent random seeds. **Table *2*** presents the reported operator errors correspond to the mean and standard deviation across these runs. The model consistently achieved a relative $L_2$ operator error in the order of $10^{-3}$, with the best performance reaching $2.97 \times 10^{-3}$. The remarkably low standard deviation in the final error metrics across different seeds demonstrates the robustness of the balanced architecture and the effectiveness of the GELU-based dual-pathway design in approximating the fluid flow operator.

**Table 2**. Statistical performance of the proposed DeepONet architecture across five independent runs.

| Metric | Mean Value | Standard Deviation | Best Achievement |
|---|---|---|---|
| Best Operator $L_2$ Error (%) | $0.355$ | $0.03$ | $0.297$ |
| Training Time (min) | $256.65$ | $14.91$ | $238.61$ |

Similarly, for DeepONet, the results shown correspond to the model achieving the median performance across the five runs.

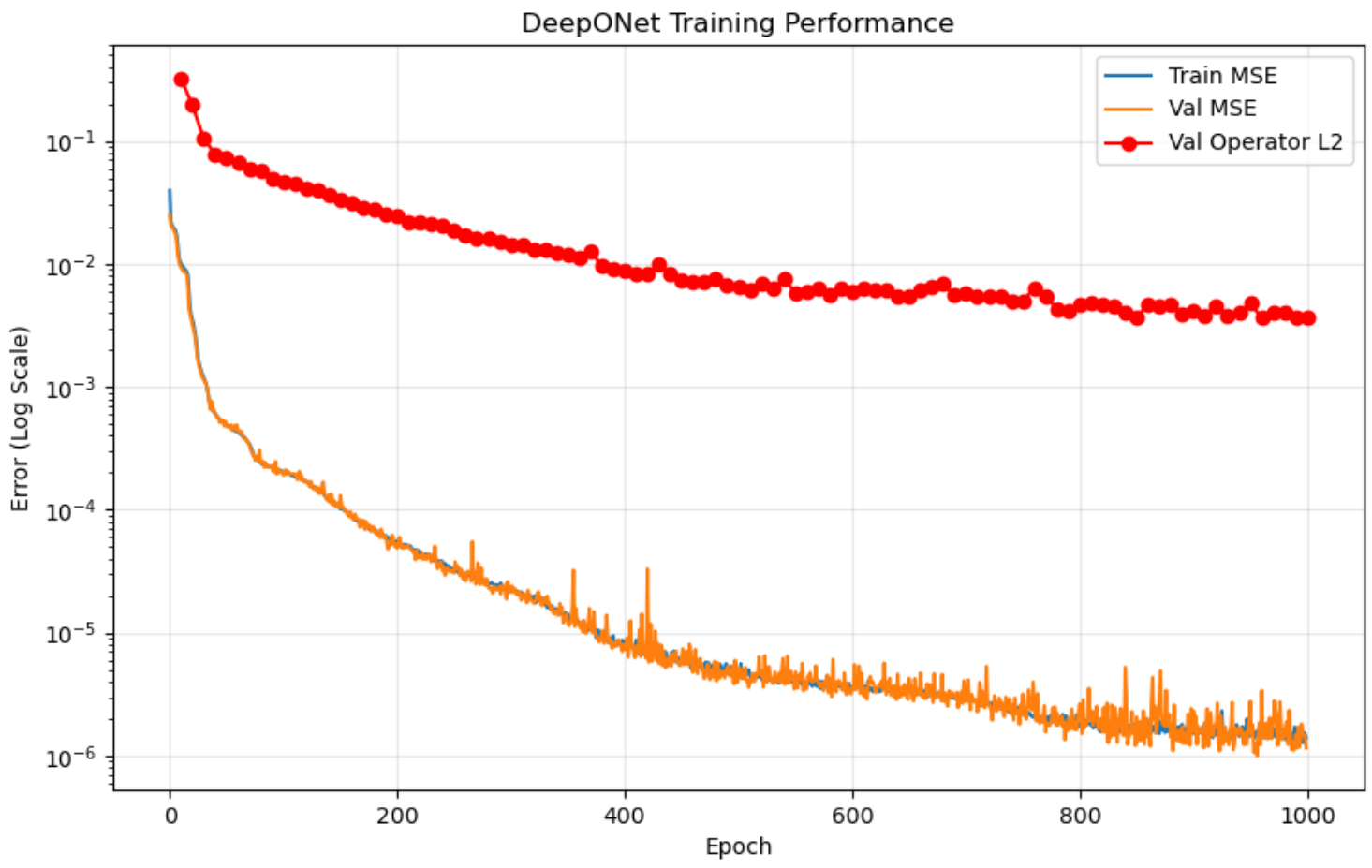

**Figure 6.** Training history of the DeepONet model showing the evolution of the mean squared error (MSE) and the relative operator $L_2$ error over epochs. Both training and validation losses exhibit rapid convergence and remain closely aligned, indicating stable learning and strong generalization.

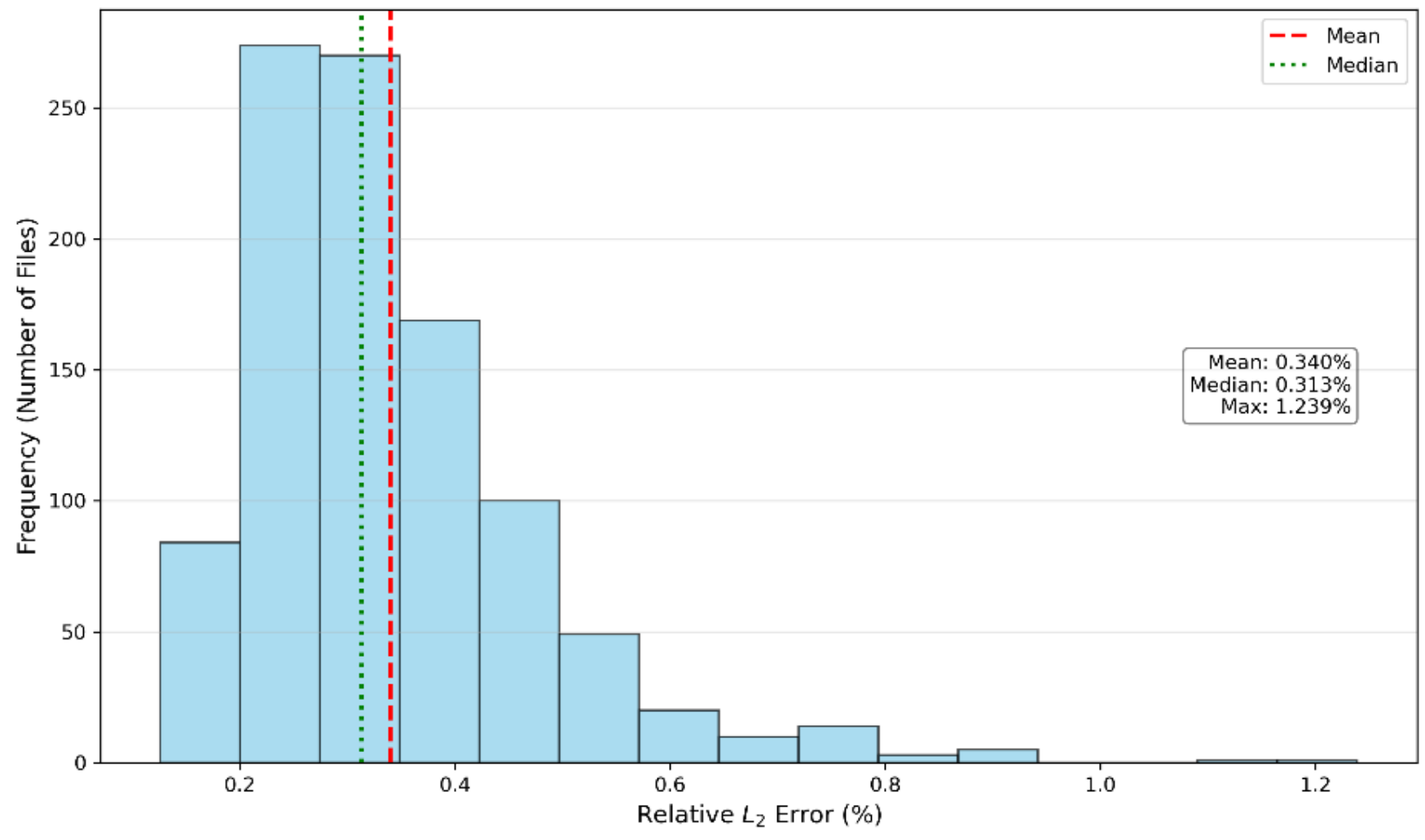

**Figure 7***.* Distribution of the relative $L_2$ error for DeepONet predictions on the unseen test set**.**

The training performance of the DeepONet architecture is illustrated in **Figure *6***, displaying the evolution of the Mean Squared Error (MSE) and the Operator $L_2$ error over $1{,}000$ epochs. Both the training and validation MSE exhibit rapid convergence during the early training stage, decreasing by several orders of magnitude from approximately $10^{-1}$ to below $10^{-4}$ within the first few hundred epochs. The close agreement between training and validation losses indicates stable learning and excellent generalization without signs of overfitting, a behavior further supported using the GELU activation function which ensures smooth gradient flow. The validation operator $L_2$ error also decreases steadily and stabilizes around $3.6 \times 10^{-3}$ ($\approx$ $0.36\%$), confirming that the network successfully learns the underlying solution operator. In the later stages of training (approximately $600 - 1{,}000$ epochs), the loss curves reach a refinement phase where the MSE approaches $10^{-6}$, with minor oscillations typical of stochastic optimization. Overall, these results demonstrate efficient and stable convergence of the DeepONet architecture.

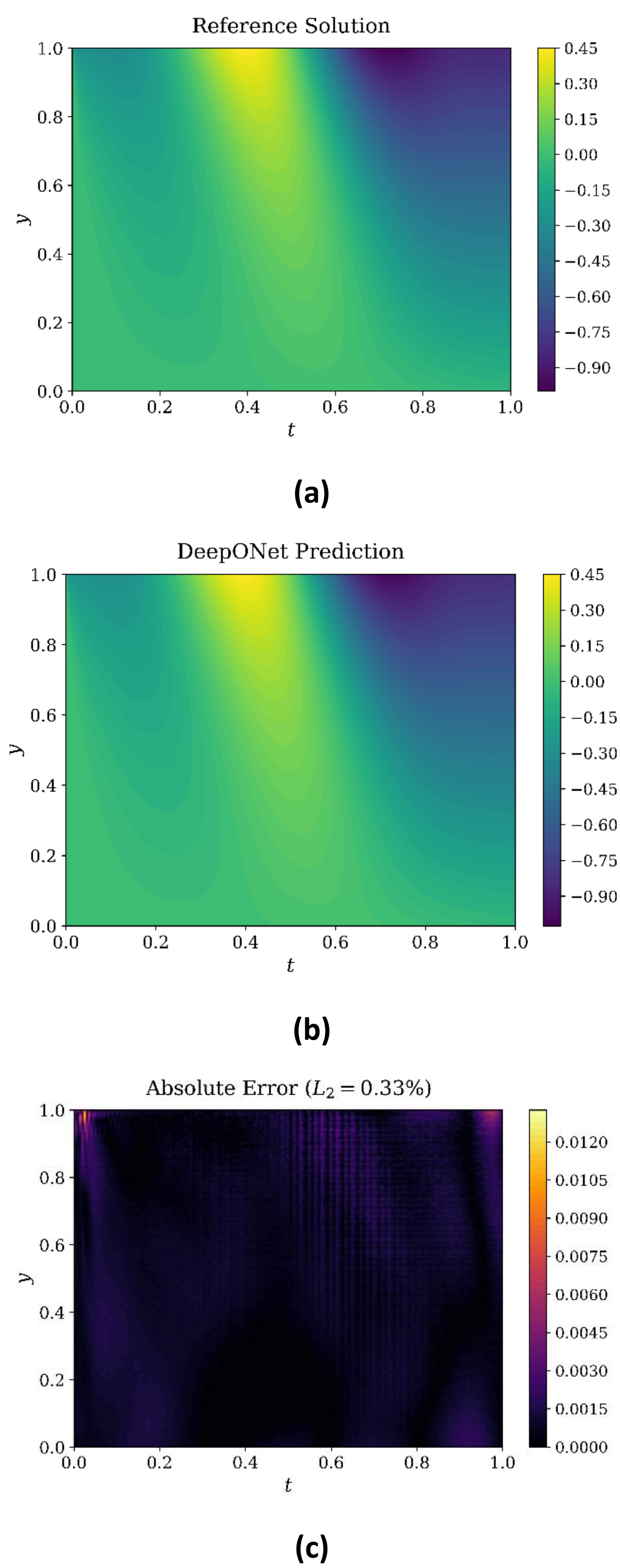

**Figure 8.** Results for a representative sample from the unseen test dataset with $B = 11.16$ and $\Lambda_s = 0.12$: (a) reference solution; (b) DeepONet prediction; (c) absolute $L_2$ error.

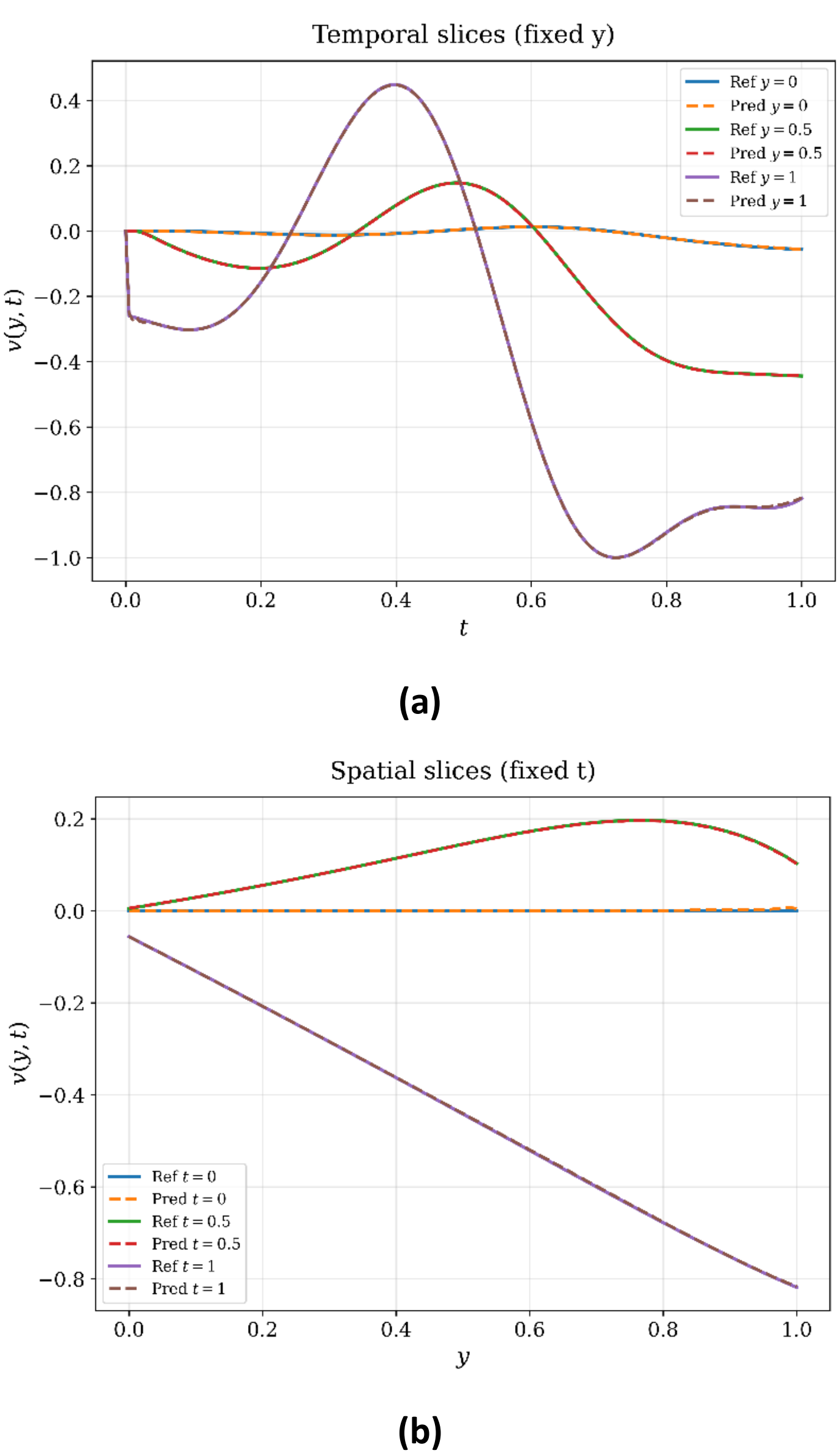

**Figure 9.** Comparison of the DeepONet predictions (dashed lines) with reference solutions (solid lines) with $B = 11.16$ and $\Lambda_s = 0.12$ : (a) Temporal slices $v(y, t)$ at fixed spatial locations; (b) Spatial slices at fixed time instances.

The predictive performance of the proposed DeepONet was validated against the unseen test dataset to assess its generalization capability. **Figure *7*** presents the distribution of the relative $L_2$ error for these samples. The model exhibits high fidelity, achieving a mean relative error of $0.340\%$ and a median of $0.313\%$. The error distribution is tightly clustered, with most cases falling below $0.5\%$. Even at the tail of the distribution, the maximum recorded error is only $1.239\%$, indicating that the model provides consistent and reliable predictions across the entire parametric range of the test set.

To further evaluate the model's performance, a representative sample from the unseen test set was selected for detailed visualization. **Figure *8*** presents a comparison between the reference solution and the DeepONet prediction for a representative sample from the unseen test dataset. The predicted solution closely matches the reference across the entire spatio-temporal domain, reproducing both the amplitude and the overall structure of the solution. The relative $L_2$ error is approximately $0.33\%$, indicating high predictive accuracy and good generalization capability of the learned operator. The absolute error map shows that the error remains very small throughout most of the domain, with slightly larger values appearing in regions where the solution exhibits stronger gradients. Similar behavior was reported by Wang et al. [32], who observed that neural operator approximations tend to exhibit higher local errors in regions with steep gradients and rapidly varying solution features. Temporal slices (fixed $y$) and spatial slices (fixed $t$) (see **Figure *9***) further confirm the excellent agreement between the predicted and reference solutions, demonstrating that DeepONet accurately captures both the temporal dynamics and the spatial variation of the solution.

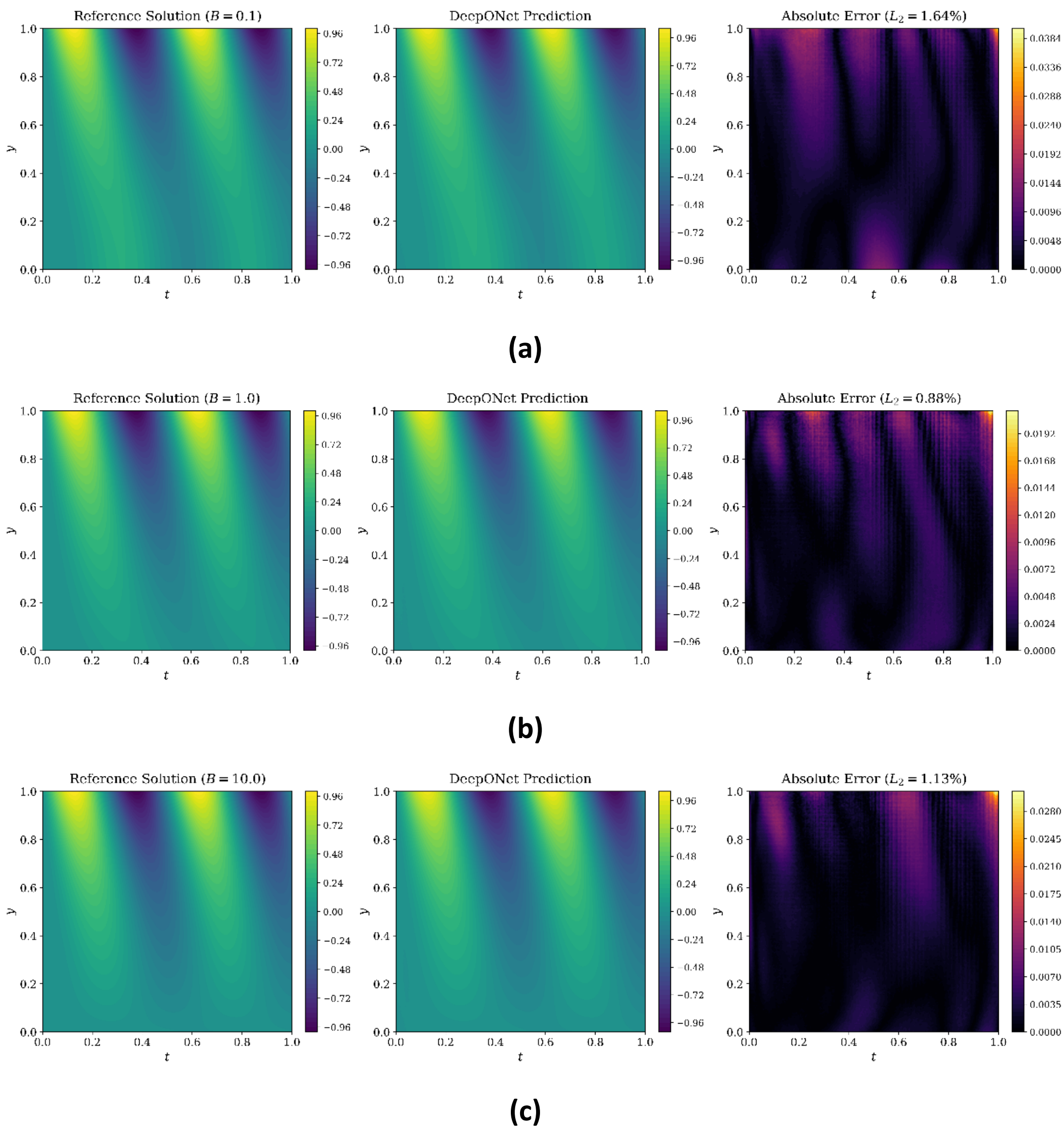

**Figure 10.** Generalization capability of DeepONet on unseen harmonic signals ($f(t) = \sin(4\pi t)$). Comparative heatmaps for (a) $B = 0.1$, (b) $B = 1$, (c) $B = 10$ with $\Lambda_s = 0.5$.

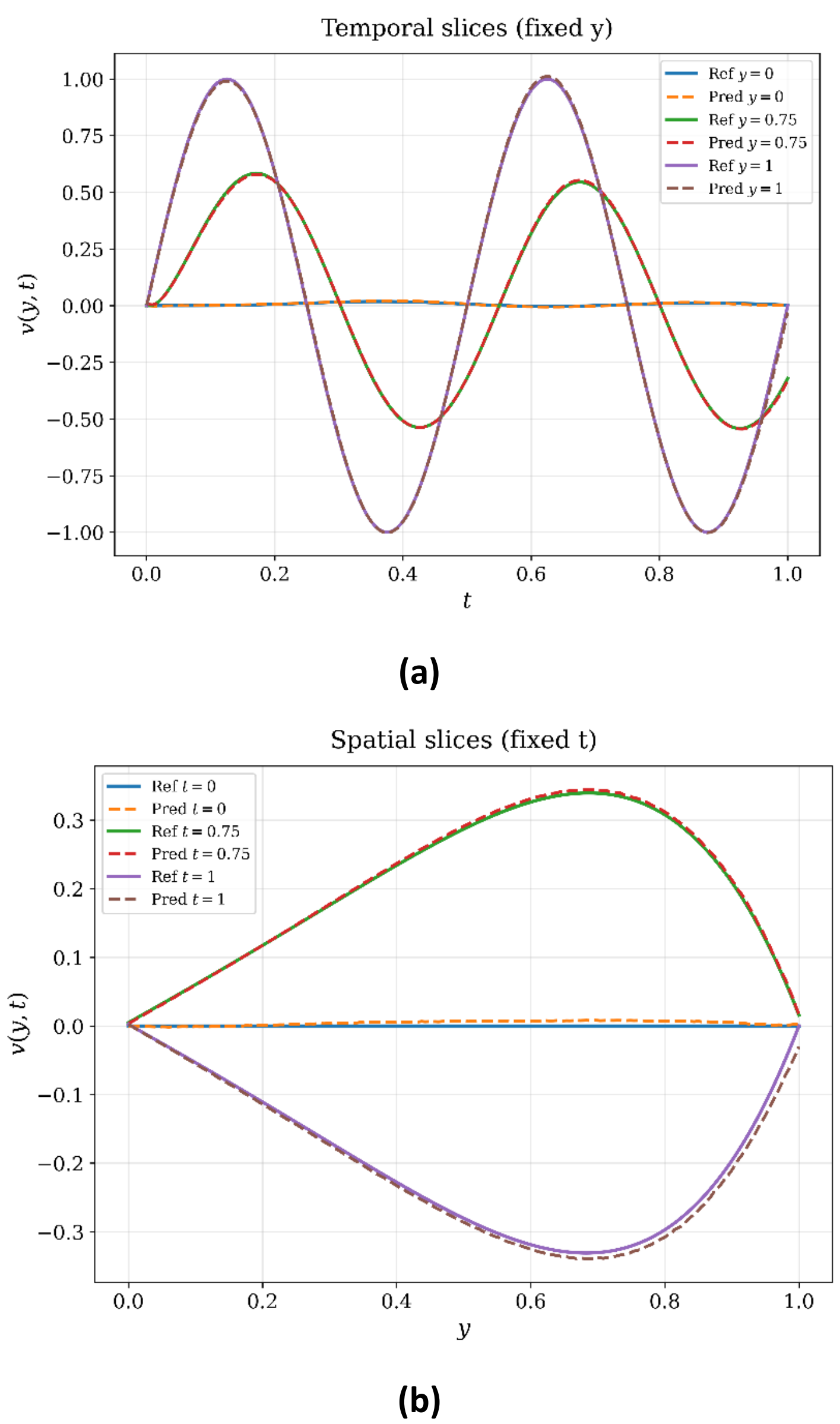

**Figure 11.** Results for $B = 10$ *and* $\Lambda_s = 0.5$ demonstrating near-perfect alignment between the reference solution (solid lines) and DeepONet predictions (dashed lines): (a) temporal slices; (b) spatial slices.

**Generalization Performance on Out-of-Distribution (OOD) Data**

To evaluate the extrapolation capabilities and the robustness of the trained DeepONet, the model was tested against a purely out-of-distribution (OOD) input signal, defined as $f(t) = \sin(4\pi t)$. This specific signal was intentionally excluded from both the training and testing datasets to serve as a rigorous benchmark for the network's generalization performance.

**Figure 10** presents the comparative heatmaps for $B = 0.1$, $1$ and $10$ with $\Lambda_s = 0.5$. The DeepONet predictions exhibit remarkable fidelity to the numerical reference solution across the entire parametric range. The relative $L_2$ error remains exceptionally low, ranging from $0.88\%$ to $1.64\%$. Notably, the absolute error maps indicate that the model maintains its accuracy even for $B = 0.1$, where the velocity gradients near the boundaries are most pronounced due to the slip conditions.

The structural integrity of the predicted velocity field is further confirmed by the spatial and temporal slices (Figure 11). In the temporal slices (fixed $y$), the DeepONet captures the frequency and phase of the oscillating flow with high precision, accurately following the $4\pi t$ harmonic profile. Similarly, the spatial slices (fixed $t$) demonstrate an almost perfect overlap between the predicted profiles and the reference solution (dashed vs. solid lines). The model successfully reconstructs the boundary layer dynamics at both $y = 0$ and 1, proving that it has internalized the underlying physical constraints and the governing operator of the oscillating fluid system.

These results emphasize that the proposed architecture does not merely perform high-dimensional interpolation but has effectively learned the continuous solution operator, allowing for reliable real-time predictions in unseen physical scenarios.

### 3.3 Comparative Analysis: PINN vs. DeepONet

In Table 3, we compare the performance of the PINN and DeepONet models developed in the previous sections against the classical Crank-Nicolson scheme. All timing measurements were performed on a Google Colab environment equipped with an NVIDIA T4 GPU, using 100 independent runs with 20 random test cases per run to ensure statistical significance.

The comparison highlights a fundamental trade-off between prediction accuracy and computational efficiency. In terms of predictive accuracy, the PINN achieves a superior relative $L_2$ error of $0.083\%$, outperforming the DeepONet's error of $0.878\%$. This high precision is expected, as PINNs function as mesh-free solvers optimized for a specific set of parameters and forcing functions, directly enforcing the underlying partial differential equations (PDEs) during training.

**Table 3.** Performance comparisons between Numerical Solver, PINN, and DeepONet.

| Method | Hardware | Rel. $L_2$ Error (%) | Inference Time (ms) (Mean $\pm$ Std) | Speedup |
|---|---|---|---|---|
| **CN** | CPU | *Reference* | $109.4 \pm 1.85$ | $1 \times$ |
| **PINN** | GPU | 0.083 | $0.61 \pm 0.076$ | $\sim 17.93 \times$ |
| **DeepONet** | GPU | 0.08783 | $0.02 \pm 0.084$ | $\sim 540.59 \times$ |

However, this accuracy comes at a significant computational cost when multiple problem instances are considered. As shown in **Table *3***, the DeepONet demonstrates an average inference time of just $0.02 \pm 0.084\ ms$ per solution, making it approximately $30.5 \times$ faster than the PINN's inference ($0.61 \pm 0.076\ ms$) and over $540.59 \times$ faster than the classical CN scheme ($109.4 \pm 1.85\ ms$). More importantly, it serves as a universal operator learner; once trained on a diverse dataset of input functions $f(t)$ and slip parameters ($\Lambda_s, B$), it provides near-instantaneous predictions for any unseen combination within its parametric space without requiring additional training or re-solving of the governing equations. In contrast, the PINN requires a time-consuming retraining process for each new scenario, limiting its utility in real-time applications or large-scale optimization loops. Consequently, while PINNs can achieve higher accuracy for individual problem parameters, DeepONet provides significantly faster inference once trained, making it particularly suitable for applications that require rapid evaluation of the solution across many parameter configurations.

## 4. Conclusions

We have investigated the performance of two prominent Scientific Machine Learning (SciML) frameworks, i.e., Physics-Informed Neural Networks (PINNs) and data-driven Deep Operator Networks (DeepONets), in solving the plane Newtonian Couette flow with dynamic wall slip. The comparative analysis revealed that both architectures effectively bridge the gap between classical numerical methods and neural-based solvers, albeit with different strengths.

The PINN framework achieves high accuracy, yielding a relative $L_2$ error of $0.083\%$. However, its inherent limitation lies in its instance-specific nature, requiring significant computational resources for retraining whenever physical parameters or boundary conditions change. In contrast, the developed data-driven DeepONet, once trained on high-fidelity numerical data, successfully learned the underlying solution operator. It demonstrated excellent generalization capabilities, maintaining sub-percentage errors even for unseen and out-of-distribution harmonic signals.

A significant finding of this study is the dramatic improvement in computational efficiency offered by DeepONet. With an inference time of approximately $109.4\, ms$ per solution, it achieved a speedup factor of $540.59\times$ over the Crank-Nicolson method and outperformed the PINN's inference by an order of magnitude. These results indicate that while PINNs remain superior for high-precision, single-instance verification, the DeepONet's ability to provide instantaneous, parametric-aware predictions makes it the ideal candidate for real-time monitoring, digital twins, and industrial design optimization.

Future work will explore the integration of physics-informed losses within the DeepONet architecture to further enhance its accuracy while maintaining its operational speed and the extension of the method to higher-dimensional flows, e.g., transient Poiseuille flow in a square channel [15], reducing data requirements through hybrid training strategies.

Overall, this study has demonstrated the complementary roles of physics-informed and operator-learning approaches in scientific machine learning, providing a pathway toward efficient and scalable modeling of complex flows.